# Self-focusing of CW Laser Beam with Variable Radius in Rubidium Atomic Vapor

V.A. Sautenkov[a,b,*], M.N. Shneider[c], S.A. Saakyan[a], E.V. Vilshanskaya[a,d], D.A. Murashkin[a], I.D. Arshinova[a], B.V. Zelener[a], B.B. Zelener[a,d,e]

[a]*Joint Institute for High Temperatures, Russian Academy of Sciences, Moscow 125412, Russia*
[b]*P. N. Lebedev Physical Institute, Russian Academy of Sciences, Moscow 119991, Russia*
[c]*Department of Mechanical and Aerospace Engineering, Prinseton University, Princeton, NJ 08544, USA*
[d]*Moscow Power Engineering Institute, Moscow 111250, Russia*
[e]*Moscow Engineering Physics Institute, Moscow 115409, Russia*

**Abstract**

Self-focusing of a cw laser beam in rubidium atomic vapor was studied. The beam power and beam spot size at entrance of a glass cell with the rubidium vapor were variable parameters. A steep grow of the threshold power of self-focusing for the small beam radii (< 30 μm) was observed. Our experimental data are in an agreement with the theoretical results published by Semak and Shneider in 2013. Additional experiments in resonance and transparent media are suggested in order to check and extend our measurements.

*Keywords:* Nonlinear optics; Self-focusing; Atomic spectroscopy.

## 1. Introduction

Effects of self-focusing and filamentations of optical beams are very interesting and important phenomena in nonlinear optics [1]. The first publication about possibility to observe self-focusing electromagnetic emission in non-linear media was appeared in 1962 [2]. A little bit later a theoretical model was developed [3,4]. These nonlinear effects were observed in transparent media [1] and also in absorptive resonance gases like atomic vapors [5,6].

Propagation of the optical beam in a linear medium and in nonlinear media is different. Intensity distribution of a Gaussian optical beam in free space (vacuum) depends on the axial and transverse coordinates $z$ and $r$ [7] and described by equation

$$I(r,z) = \frac{2P}{\pi w(z)^2} \exp\left(-\frac{2r^2}{w(z)^2}\right), \tag{1}$$

where $P$ is the power of the optical beam, $w(z)$ is the radius of the beam. The Gaussian beam is contracted to minimum sport size $w_0$ at the beam waist. The radius of the beam $w(z)$ depends on the distance $z$ from the waist ($w(0) = w_0$). The expansion law for the optical beam radius $w(z)$ is expressed as

\* Corresponding author.
*E-mail address:* vsautenkov@gmail.com.

$$w(z)^2 = w_0^2 \left(1 + \frac{\lambda z}{\pi w_0^2}\right)^2. \tag{2}$$

When a suitable thin lens installed in the plane $z = 0$ a Gaussian beam with a beam spot size $w_{01}(0)$ can be converted to a waist with a sport size $w_{02}(f)$. When confocal parameter $\pi w_{01}(0)^2/\lambda$ is much more than the focal length $f$, the next relation can be used for estimation $w_{02}(f)$ [8]

$$w_{02}(f) = \frac{\lambda}{\pi w_{01}(0)} f. \tag{3}$$

The lenses with different $f$ can be used to control a spot size of entrance beam into a nonlinear medium. In nonlinear medium the optical beam can induce intensity dependent lenses

$$n(r, z, I) = n_0 + n_2 I(r, z). \tag{4}$$

In such medium the expansion of the optical beam is change to compare with the free space. It modified by lens-like properties of the medium.

In model of [3] at the level of a critical power $P_{cr}$ the diffraction expansion of the optical beam is compensated by optically induced lens-like properties of the nonlinear medium. A simple formula for the critical power $P_{cr}$ was obtained in [3]

$$P_{cr} = \xi \left(\frac{\lambda^2}{4\pi n_0 n_2}\right), \tag{5}$$

where the dimensionless factor $\xi$ is order of 2. According this formula the critical power is independent on the optical beam sport size. The formula was applied for estimation of critical power in a set of experimental research works [1].

The detailed theoretical consideration of Gaussian beam propagation in the non-linear medium was performed in [9]. The threshold power of the laser beam self-focusing was defined as the power when the wave front becomes flat after propagating some distance through the nonlinear media. For small radii of the optical beams ($w_0 < 50$ μm) a steep grow of the threshold was predicted and discussed. Recently a preliminary experimental confirmation of these theoretical results was published [10]. The laser beam was contracted by a lens at the entrance window of a cell with rubidium atomic vapor. Variation of the self-focusing threshold for different focus length of the lens was observed.

In present paper the propagation of the laser beam with the variable radius in resonance gas was investigated. The laser beam was focused on a glass cell with rubidium atomic vapor in such way that the minimal radius of the beam was formed at an interface glass/vapor. The dependence of the self-focusing threshold power on the beam's radius was studied. More accurate measurements allowed us to compare our experimental data with theoretical results of [3,9].

## 2. Experimental arrangement

The experiment was performed with rubidium atomic vapor at $D_2$ line (wavelength $\lambda = 780$ nm). A natural abundance of [85]Rb (72.2 %) and [87]Rb (28%) was used. Scheme of our experimental setup is presented in Figure 1. The cw external cavity diode laser (ECDL) was used as a sourse of a narrow band emission (laser linewidth < 1 MHz). The laser frequency can be tuned near $D_2$ line of rubidium ($\lambda = 780$ nm). By the first beamsplitter (BS1) a small part of the output laser emission was sent to a wavemeter.

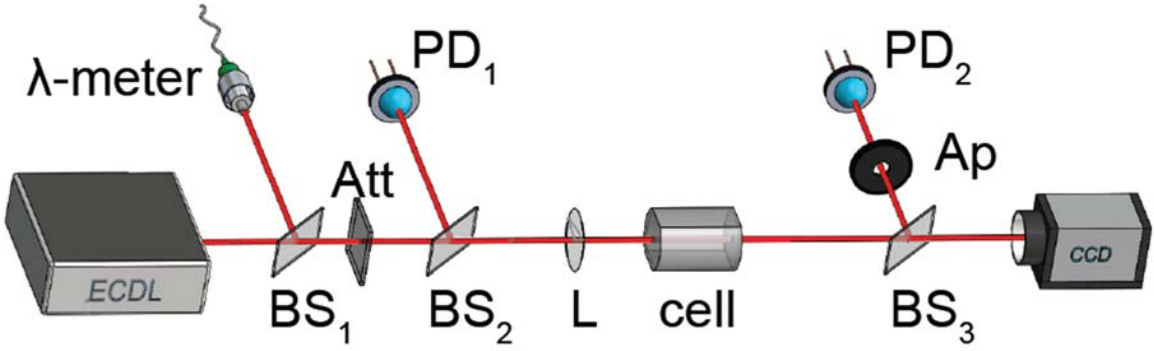

**Fig. 1.** Scheme of experimental setup: ECDL is external cavity diode laser, BS1, BS2, BS3 are beam splitters, λ-meter is wavemeter Angstrom, Att is variable attenuator, L is lens, cell is glass cell with hot rubidium vapor, CCD is charge-coupled device, PD1 and PD2 are photo detectors, Ap is variable aperture.

The measurements were made with a glass cell which contained a drop of rubidium metal. The length of the cell is 9 cm. The cell with rubidium vapor can be heated. The temperature of the cell was measured by thermocouples. The atomic number density $N$ of rubidium vapor was defined by the temperature of coldest sport of the cell [11]. The absorption spectrum of the rubidium vapor at room temperature ($N \approx 10^{11}$ cm$^{-3}$) is shown in Figure 2. The Doppler width $\Delta\omega_D$ of the atomic transitions is the order of $2\pi \times 0.5$ GHz. It is more than spectral intervals between sublevels of the hyperfine structure (hfs) of the excited states $5P_{3/2}$ in $^{85}$R and $^{87}$Rb atoms.

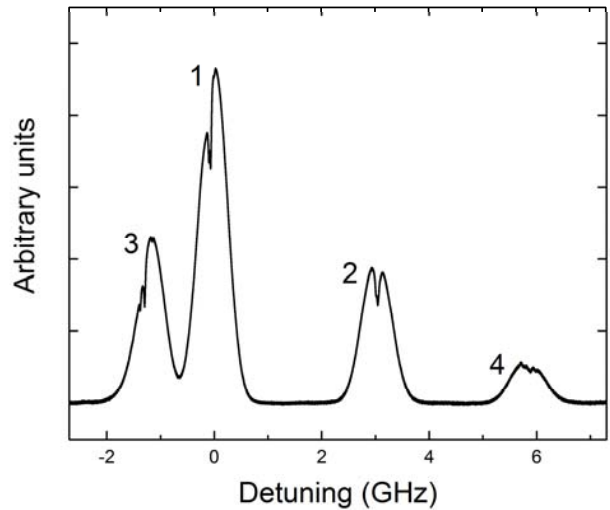

**Fig. 2.** Absorption spectrum of the rubidium vapor at a room temperature. Resonances 1 and 2 represent sets of the hfs transitions in $^{85}$Rb atoms. Resonances 3 and 4 2 represent sets of the hfs transitions in $^{87}$Rb.

The absorption spectrum consist four spectral resonances. Each resonance is a combination of three hfs transitions. The natural width of the hfs transitions $\gamma_{nat}/2\pi$ is equal to 6 MHz. Resonances 1 and 2 represent sets of the hfs transitions $5S_{1/2}(F = 3) \rightarrow 5P_{3/2}(F' = 2, 3, 4)$ and $5S_{1/2}(F = 2) \rightarrow 5P_{3/2}(F' = 1, 2, 3)$ in [85]Rb atoms. Resonances 3 and 4 represent sets of the hfs transitions $5S_{1/2}(F = 2) \rightarrow 5P_{3/2}(F' = 1, 2, 3)$ and $5S_{1/2}(F = 1) \rightarrow 5P_{3/2}(F' = 0, 1, 2)$ in [87]Rb atoms.

The self-focusing was investigated with the rubidium atomic number density $N$ of $5.7 \times 10^{13}$ cm[-3]. Such density was chosen in order to obtain a reasonably large nonlinear part of refraction index n$_2$ [12] and a reasonably small self-broadening $\Gamma$. It is described by a relation $\Gamma = KN$ [13]. For the hfs components of D$_2$-lines of [85]R and [87]Rb atoms the factor $K$ is of $2\pi \times 10^7$ sec[-1] cm[3] and under experimental conditions self-broadening $\Gamma$ is of $2\pi \times 5.7$ MHz [13]. In our case self-broadening is less than the natural width. In the atomic density region, where relation $\Gamma \leq \gamma_{nat}$ is satisfied, a possible dependence of the spectral width the optical saturation can be neglected [14,15]. At high atomic densities, where $\Gamma \gg \gamma_{nat}$, self-broadening can depend on the level of the atomic excitation [15-17]. In our experiment the robust self-focusing was observed at frequency detuning near 9 GHz from hfs transition $5S_{1/2}(F = 3) \rightarrow 5P_{3/2}(F' = 4)$ to the blue wing of the D$_2$-line. As the next step we would like to estimate the nonlinear refraction index for our experimental condition. The refraction index is can be written as $n^2 = (1 + \chi)$ by using relations for the atomic gas susceptibility from textbook [12]. For large detuning of laser frequency $\omega_{las}$ from transition frequency $\omega_{ab}$ when detuning $\Delta$ is much more than Doppler broadening and homogeneous width $1/T_2$, the contribution of the thermal motion of atoms to the susceptibility can be neglected. In this case an expression for the nonlinear susceptibility from [12] for a two-level atomic system can be used

$$\chi^{(3)} = \frac{\alpha_0(0)}{3\omega_{ba}/c} \left[\frac{\Delta T_2 - i}{(1+\Delta^2 T_2^2)^2}\right] \frac{2\varepsilon_0 c}{I_s^0}, \tag{6}$$

where $\alpha_0(0)$ is unsaturated, line-center absorption coefficient, $\varepsilon_0$ is dielectric constant for vacuum, $I_s^0$ is saturation intensity of the atomic transition, $\Delta$ is detuning $(\omega_{las} - \omega_j)$ and the width $1/T_2$ is $(\gamma_{nat} + \Gamma)$.

By using the equations (6) and the next equation 7

$$n_2 = \frac{3}{4n_0^2 \varepsilon_0 c} \chi^{(3)}, \tag{7}$$

it is possible to estimate $n_2$.

For our experimental parameters we wrote a simplified expression $n_2 \approx \Sigma \lambda \alpha_i (32\pi I_{(j)s})^{-1} (\gamma_{nat} + \Gamma)^3 (\omega_{las} - \omega_j)^{-3}$. Here the subscript $(j)$ indicates that this parameter is related to one of selected atomic transitions which are taken into account. The estimation of $n_2$ is lying between 10[-8] cm[2]/W and 10[-7] cm[2]/W.

## 3. Experimental results and discussions

In order to control the spot size of the optical beam on the entrance window of the cell a set of lenses was used. The intensity distribution of the laser beam before the lens L was close to the Gaussian function (beam radius 0.8 mm). The laser beam was focused on the first entrance window of the cell in such way that the minimum radius $w_0$ of the laser beam (the beam waist) was coincided with the interface glass/vapor. A set of lenses with focal length $f$ from 30 cm to 6 cm was allowed us to vary the radius $w_0$ from 0.1 mm to $2 \times 10^{-2}$ mm. The intensity distribution in the plane corresponding to the interface glass/vapor was recorded by CCD. For the lens L with $f = 15$ cm. The intensity profile, obtained by using the lens with $f = 15$ cm, is shown in Figure 3.

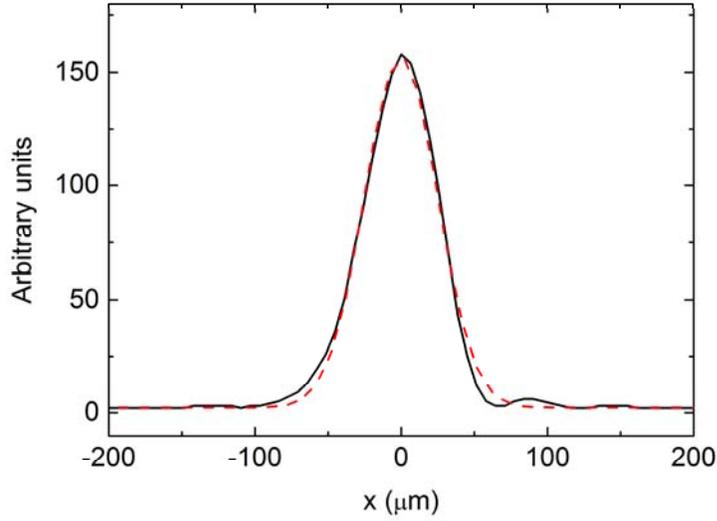

**Fig. 3.** Intensity profile of input beam along horizontal axis $x$ for lens with $f = 15$ cm. The red dashed curve is result of the fit by Gaussian function with $w_0 = 50$ μm.

In Figure 3 the black solid curve presents experimental results, red dashed curve is result of the fit by the Gaussian function $F(x) = A exp(-2x^2/w_0^2)$. The fitting parameter $w_0$ is of 50(0.3) μm. The intensity distributions of the optical beam after passing through the rubidium vapor were recorded by CCD. A sensitive area of CCD was $8.6 \times 8.67$ mm². The distance from the output window of the cell to the imaging device was of 15 cm. By analyzing of the recorded patterns for different input optical powers it was possible to select a power region where a power independent beam profile was observed (the self-trapping propagation). Outside this region the power dependent beam profiles were observed. We defined a threshold power $P_{SF}$ as a low power boundary of the stable power region. Our definition is close to the definition of the threshold power for self-focusing in [9].

The images for three input optical powers are presented in Figure 4. A lens with focus length $f = 15$ cm was used to prepare the input beam. The intensity distribution of the beam after passing the cell was practically the same in the power region from 3.45 mW to 4.1 mW. The threshold power $P_{SF}$ determined as 3.45 mW. The image in Figure 4 (a) was obtained in a low power region where $P < P_{SF}$. The image in Figure 4(b) was obtained inside the stable power region (3.45 mW to 4.1 mW). The image in Figure 4(c) obtained in a high power region where $P > 4.1$ mW.

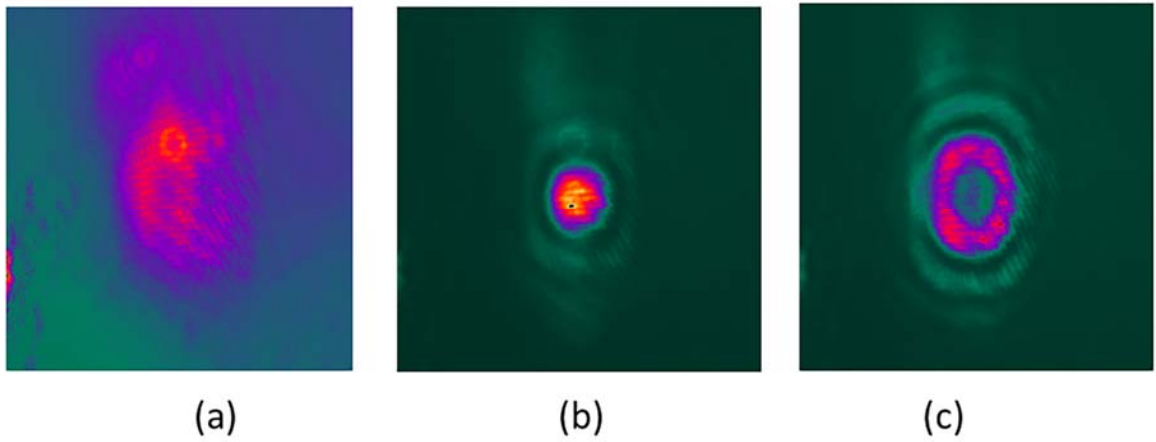

**Fig. 4.** Intensity distribution of output beam at the distance 15 cm from the second cell window. The images were recorded using the lenses L with $f = 15$ cm. (a) $P = 1.2$ mW, (b) $P = 3.8$ mW, (c) $P = 5.1$ mW.

A non-regular structure of pattern in Figure 4(a) is related with optically induced lens-like properties of the rubidium vapor together with scattering of the laser beam on surfaces of the cell windows. A symmetrical bright spot in Figure 4(b) corresponds to the self-trapping propagation of the laser beam in rubidium vapor similar to observations in [18]. The cross-section of the intensity distribution is shown in Figure 5 for power 3.6 mW.

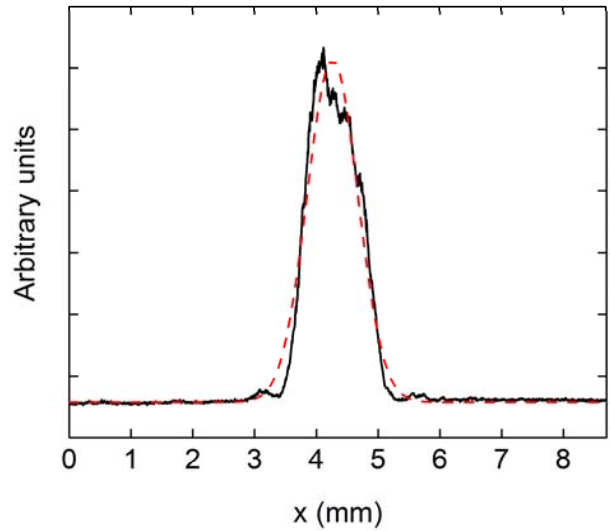

**Fig. 5.** Intensity profile of output beam along the horizontal axis $x$. The black said curve is result of measurements, the red dashed curve is result of the fit by Gaussian function with $w = 0.82$ mm.

In Figure 5 the black solid curve is result of measurements, the red dashed curve is result of the fit by Gaussian function $F(x) = Aexp(-2(x - x_c)^2/w^2)$. The fitting parameters w and $x_c$ are 0.82 mm and 4.26 mm respectively. Uncertainties are less than the 0.01 mm. One can see that the profile is quite well described by the Gaussian function. We measured the threshold power $P_{SF}$ for lenses with different focal lengths f by analyzing the images of the intensity distribution for the optical beam after the cell. In addition the ration of the power of the beam passed through an aperture with diameter $D$ and the power of the beam before the passing of the aperture was measured. The diameter $D$ was fixed as the two radii of the laser beam measured in the self-trapping region (d). For example, for lens with $f = 15$ cm the selected diameter is of 0.1 mm. Both approaches gave the similar values of the self-trapping threshold power $P_{SF}$ with experimental uncertainty near 7%. The self-trapping threshold powers for different sizes of the input beam are presented in Figure 6.

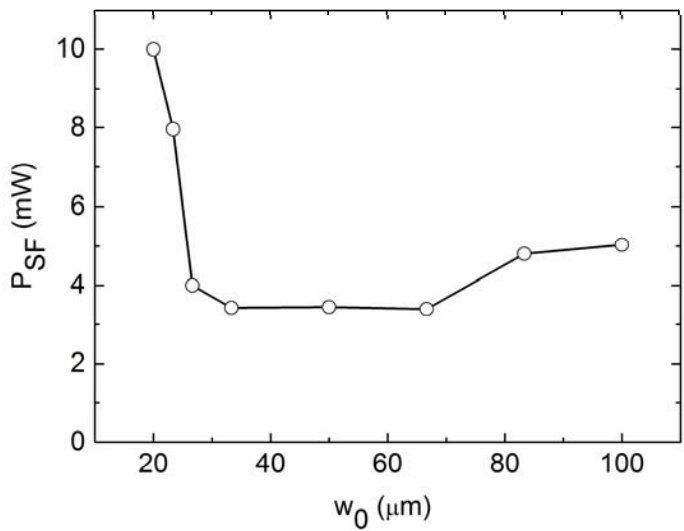

**Fig. 6.** Threshold power $P_{SF}$ vs the radius of the input beam $w_0$.

For the radii less than 30 µm the threshold power was increased fast. Growing of the threshold power for small radii of the laser beams can be associated with the diffusion of the excited atoms from the interaction area $\pi w_0^2$. The thermal velocity of atoms in the hot vapor is expressed as $v_{th} = (2k_BT/m)^{1/2}$, where $T$ is the temperature of the vapor, $k_B$ is Boltzmann constant and $m$ is mass of individual atom. For the vapor temperature 400 K the estimated thermal velocity $v_{th}$ is of $2.8 \cdot 10^2$ m/sec. The free mean path of the atomic excitation $l_{exc} \approx v_{th}/(\gamma_{nat} + \Gamma)$ is near 4 µm. This value corresponds approximately to 20% of the minimal radius $w_0$ in Figure 6. It is clear, that the diffusion of the excitation can not explain the steep variation of the threshold power for the small radii. The diffusion of the excitation can induce only modest changes of the threshold. The influence of the diffusion can be effectively reduced at higher atomic number densities. Our experimental data are in the reasonable agreement with the results of detailed calculations in [9]. By using the estimated values for $n$ and $n_2$ and the equation (5) it is possible to evaluate the critical power $P_{cr}$ as $10^{-1}$ - $10^{-2}$ W. The estimated $P_{cr}$ is more than the average threshold power at the flat part of the curve in Figure 6.

## 4. Conclusions

In the presented work the self-focusing of the laser beam in the glass cell with the hot rubidium vapor were studied. The beam power and beam radius were variable parameters. The self-focusing and self-trapping of the laser beam in the rubidium vapor were observed. The threshold power $P_{SF}$ for self-focusing was measured for different radii of the laser beam. The experimental data are in a reasonable agreement with the theoretical results of the theoretical work [9]. In the presented paper we defined the threshold power for self-focusing which is close to definition of the threshold power for self-focusing in theoretical work [9]. The experiments were performed with a low density rubidium vapor where self- broadening of the hfs transitions was less than the natural width. Under our experimental conditions the possible dependence of the self-broadening on the optical saturation can be neglected [14,15]. It is challenge to investigate self-focusing in the dense atomic vapor where self-broadening dominate to compare with the natural width. At such densities the self- broadening can be reduced by the optical excitation [16,17,19]. New aspects of the self-focusing phenomena can be discovered. Also it is interesting to extend the research of the self-focusing threshold to transparent gases by applying intense femtosecond pulses.

**Acknowledgements**

We are thankful to M. A. Gubin, M. A. Kazaryan, P. N. Lebedev Physical Institute RAS, Moscow, Russia and V. V. Semak, Virtual Laser Application Design, LLC , USA for useful discussions and help.

This work was supported by the Russian Academy of Sciences (Basic Research Program "Investigation of Matter in Extreme States" headed by V. E. Fortov).